\documentclass[conference]{IEEEtran}
\IEEEoverridecommandlockouts
\usepackage{cite}
\usepackage{amsmath,amssymb,amsfonts}
\usepackage{algorithmic}
\usepackage{graphicx}
\usepackage{textcomp}
\usepackage{xcolor}
\usepackage{bm}
\usepackage{balance}
\usepackage{subfigure}
\usepackage{float}
\usepackage{multirow}
\def\BibTeX{{\rm B\kern-.05em{\sc i\kern-.025em b}\kern-.08em
    T\kern-.1667em\lower.7ex\hbox{E}\kern-.125emX}}
\begin{document}

\title{Statistical Analysis of Primary and Random Clusters in 318 GHz Terahertz Channels for Industrial IoT}

\author{\IEEEauthorblockN{
Siyuan Shao\IEEEauthorrefmark{1},
Peize Zhang\IEEEauthorrefmark{1},
Pekka Ky\"osti\IEEEauthorrefmark{2}\IEEEauthorrefmark{3},
Trung Q. Duong\IEEEauthorrefmark{1}\IEEEauthorrefmark{4},
Simon L. Cotton\IEEEauthorrefmark{1}
}
\IEEEauthorblockA{\IEEEauthorrefmark{1}
Centre for Wireless Innovation (CWI), Queen's University Belfast,  BT3 9DT Belfast, United Kingdom}
\IEEEauthorblockA{\IEEEauthorrefmark{2}
Centre for Wireless Communications (CWC), University of Oulu, 90570 Oulu, Finland}
\IEEEauthorblockA{\IEEEauthorrefmark{3}
Keysight Technologies Finland Oy, 90590 Oulu, Finland}
\IEEEauthorblockA{\IEEEauthorrefmark{4}
Memorial University, St. John’s, NL A1B 3X5, Canada}
\IEEEauthorblockA{sshao02@qub.ac.uk, p.zhang@qub.ac.uk, pekka.kyosti@oulu.fi, trung.q.duong@qub.ac.uk, simon.cotton@qub.ac.uk}
}


\maketitle

\begin{abstract}
The ultra-high data rates enabled by terahertz (THz) communications pave the way for the demanding requirements of industrial Internet of Things (IIoT) applications, making the investigation of THz channels in industrial environments a critical research topic. This paper presents a comprehensive statistical analysis of the propagation channel at 318\,GHz in an industrial environment. In particular, a new clustering scheme is proposed for the sparsity observed in the multipath components (MPCs) of the measured channel. Furthermore, statistical analyses are conducted separately for the group of strong reflections, defined as primary clusters, and other propagation phenomena, defined as random clusters, in a rich-scattering environment. The results demonstrate that the large-scale parameters are predominantly influenced by these strong reflections. This study provides reliable support and guidance for subsequent THz stochastic channel modeling.
\end{abstract}


\section{Introduction}
\label{sec:I}
The terahertz (THz) spectrum, offering continuous bandwidths spanning tens of gigahertz (GHz), holds significant promise for enabling ultra-reliable low-latency communications and hyper-scale connectivity \cite{akyildiz2022terahertz}, proving vital for industrial Internet of Things (IIoT) systems reliant upon the interconnection of numerous sensors, machinery, and communication devices \cite{aceto2019survey}. Higher frequencies introduce more severe path loss and, as wavelength approaches the visible spectrum, the optical properties of wave propagation become increasingly pronounced \cite{takahashi2024channel}, resulting in the characteristics of THz wireless channels differing significantly from those of millimetre-wave channels. Unlike other indoor environments such as offices and meeting rooms, industrial environments contain numerous metallic scatterers including lathes, shelves, and machinery. Therefore, a thorough understanding and accurate characterization of THz channels in industrial settings are essential for the design and deployment of IIoT communication systems.

A stochastic channel model based on ray-tracing simulations at 300\,GHz in an indoor environment was systematically studied as early as the previous decade \cite{priebe2013stochastic}, while extensive efforts in channel measurement, characterization, and modeling above 100\,GHz have progressively emerged in this decade \cite{han2022terahertz}. The Saleh-Valenzuela (S-V) model \cite{saleh1987statistical} and the geometry-based stochastic channel model (GBSCM) \cite{ii2008winner} remain the primary methodologies for channel statistical analysis \cite{de20253gpp, ju2024statistical, chen2023channel}. The sparsity of clusters and rays in sub-THz channels has been reported in several studies \cite{de20253gpp, ju2024statistical}. This sparsity can be attributed to factors such as limited angular coverage of measurement systems and the performance of multipath component (MPC) extraction and clustering algorithms. It is also consistent with the fact that higher-order reflections experience greater attenuation at higher frequencies, rendering them indistinguishable. For instance, measurements at 142\,GHz reported an average of 6 clusters and an average of 2 rays per cluster in a residential line-of-sight (LoS) scenario \cite{de20253gpp}, while only 1.8 clusters and 1.4 rays per cluster were observed in an office LoS scenario \cite{ju2021millimeter}.

This sparsity of THz channels also poses challenges for MPC extraction and clustering. The authors of \cite{ju2024statistical} proposed a novel MPC extraction and clustering method and observed 3.4 clusters and 3.9 rays per cluster in a factory LoS scenario at 142\,GHz, indicating richer scattering in industrial settings. Furthermore, due to quasi-optical propagation properties, the first few orders of reflections in sub-THz and THz channels exhibit distinct characteristics compared to higher-order MPCs, and have been analyzed separately in prior works \cite{takahashi2024channel, chen2023channel, priebe2013stochastic}. While such analysis improves modeling accuracy of the specific channel environment, it is important to balance accuracy and randomness in channel coefficient generation.

In this paper, we investigate the wireless channel in a typical factory environment at 318\,GHz and perform a statistical analysis of the measurement data to support THz channel modeling. The main contributions are summarized as follows:
\begin{itemize}
\item The KPowerMeans clustering algorithm and the Silhouette index were extended to cater for the sparsity of MPCs in THz channels.
\item Clusters are classified into primary clusters and random clusters, and the feasibility of this approach was validated through statistical analysis.
\item Parameters of primary and random clusters, including the number of clusters and rays per cluster, interarrival time, power, and angle, were analyzed separately, and a comparative study was performed.
\end{itemize}

\section{Measurement Campaign}
\label{sec:II}
\subsection{THz channel sounder}
The measurement campaign in this work was conducted using a vector network analyzer (VNA)-based channel sounder developed by the University of Oulu. At both the transmitter (Tx) and receiver (Rx), pyramidal horn antennas operating in the 220–330\,GHz frequency band are employed, together with a frequency up-converter, a down-converter, and a local oscillator (LO), enabling channel measurements over a 4\,GHz bandwidth centered at 318\,GHz. An optical fiber cable is used to transmit the LO signal from an electrical-to-optical converter to an optical-to-electrical converter, which is connected to an attenuator and a down-converter at the Rx, supporting a wide range of measurement activities.

Within the 4\,GHz bandwidth, 1001 frequency points are sampled, resulting in a delay resolution of 0.25\,ns and a maximum excess delay of 250\,ns, sufficient to distinguish and capture the MPCs of interest in the time domain. The antenna gain is 21.69\,dBi, with half-power beam widths (HPBWs) of $14.2^{\circ}$ in azimuth and $14.4^{\circ}$ in elevation at the centre frequency of 318\,GHz. Measurements were conducted using a bidirectional scanning approach, and a back-to-back calibration was performed before measurements.

\subsection{Measurement Environment}
\begin{figure}[!t]
\centerline{\includegraphics[width=3.0in]{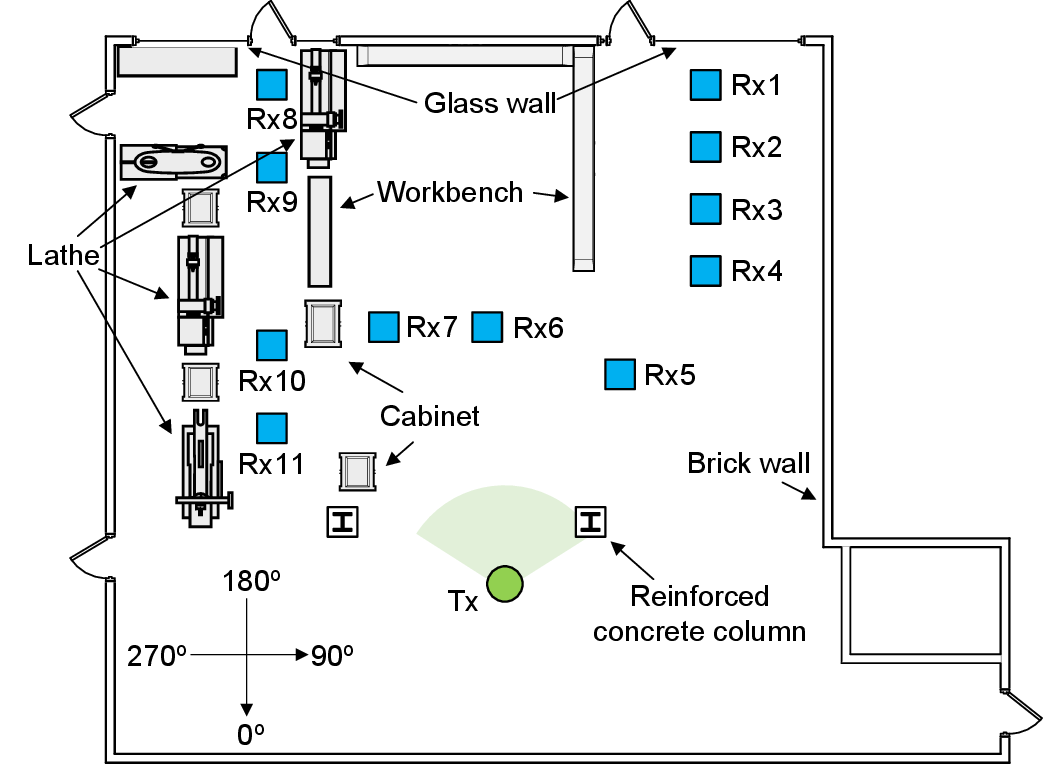}}
\caption{Layout of the factory environment.}
\label{layout}
\vspace{-6pt}
\end{figure}

The measurements were conducted in a representative factory hall containing numerous machines, lathes, and metal cabinets. In this environment, 11 Rx locations with LoS conditions were selected. As illustrated in Fig.~\ref{layout}, the Tx location is marked as a green circle and the Rx locations as blue squares on the floor plan. To approximate an omnidirectional radiation pattern in the azimuth plane through directional measurements, the Rx antenna is rotated over $360^{\circ}$ in increments of $15^{\circ}$. The Tx antenna is rotated only within the angular range of interest ([$120^{\circ}$, $210^{\circ}$]), also with increments of $15^{\circ}$. The elevation angles of both the Tx and Rx antennas are kept fixed during the measurements. The Tx antenna is slightly tilted downward, as the turrets at Tx and Rx are mounted at heights of 1.85\,m and 1.43\,m above ground, respectively. To secure a static measurement environment, the measurements were conducted overnight with no human movement, and all machinery in the test area was kept inactive.

\section{Multipath Parameter Estimation}
\label{sec:III}
\subsection{Multipath Component Extraction}
The raw data obtained from bidirectional scanning measurements correspond to the channel transfer function (CTF) $H(f,\Omega_{\text{t}},\Omega_{\text{r}})$, which can then be used to calculate the channel impulse response (CIR) $h(\tau,\Omega_{\text{t}},\Omega_{\text{r}})$ by performing an inverse fast Fourier transform on it after windowing. Subsequently, the power angular delay profile (PADP) is calculated as
\begin{equation}\label{equ:cir}
\begin{split}
P(\tau,\Omega_{\text{t}},\Omega_{\text{r}})&={\mid}h(\tau,\Omega_{\text{t}},\Omega_{\text{r}}){\mid}^2\\
&={\mid}\mathcal{F}^{-1}(H(f,\Omega_{\text{t}},\Omega_{\text{r}})W(f)){\mid}^2,
\end{split}
\end{equation}
where $\Omega_{\text{t/r}}$ denotes the direction of the Tx or Rx antenna, including azimuth angle of departure (AoD) $\varphi_{\text{t}}$, elevation angle of departure (EoD) $\phi_{\text{t}}$, azimuth angle of arrival (AoA) $\varphi_{\text{r}}$, and elevation angle of arrival (EoA) $\phi_{\text{r}}$, and $W(f)$ represents the window function. In this work, the antenna direction comprises only AoD and AoA, and a Blackman-Harris window is applied.

\begin{figure}[!t]
\centering
\subfigure[]{\includegraphics[width=2.5in]{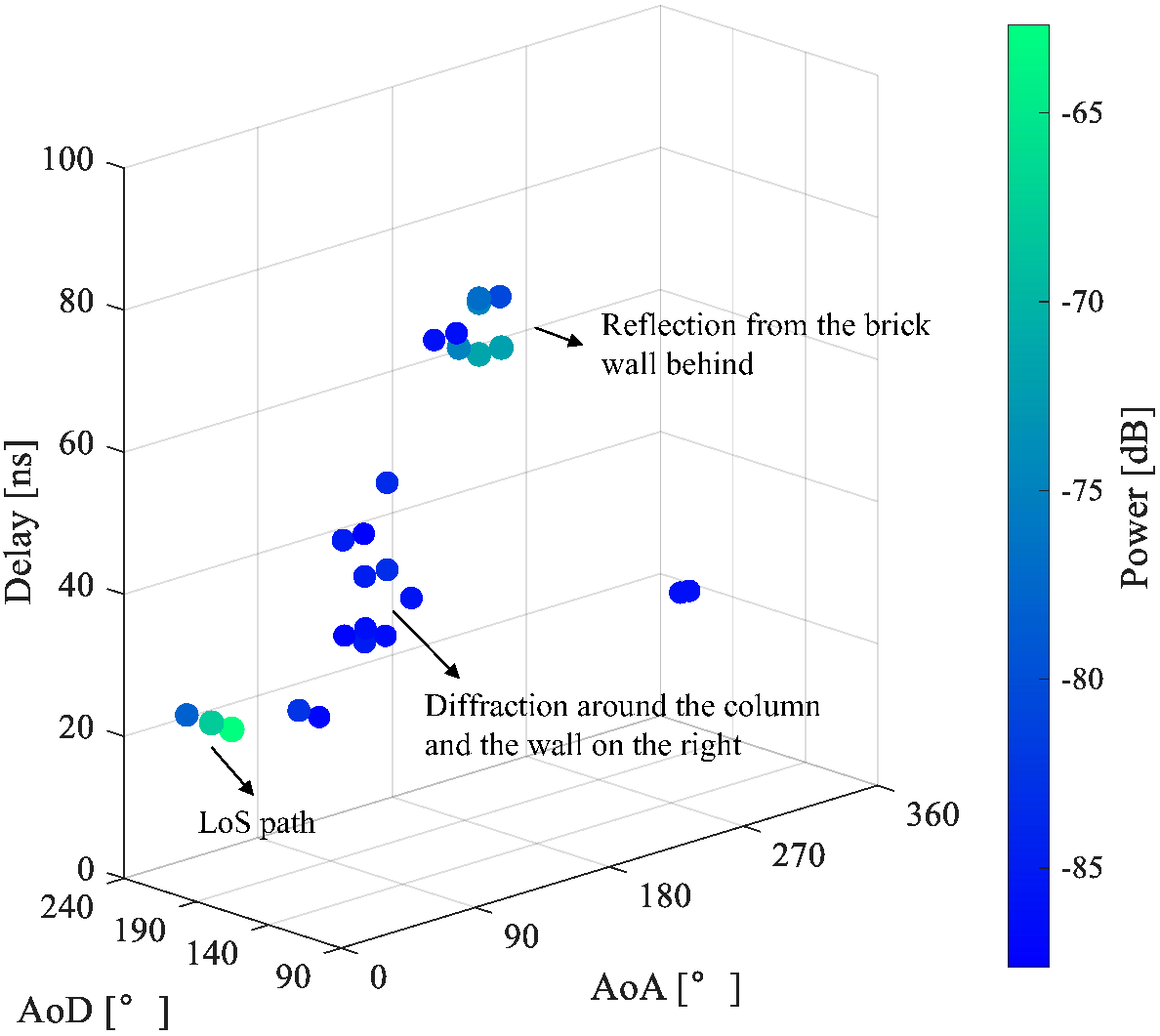}
\label{fig:PADP}}
\hspace{6pt}\\   
\subfigure[]{\includegraphics[width=3.0in]{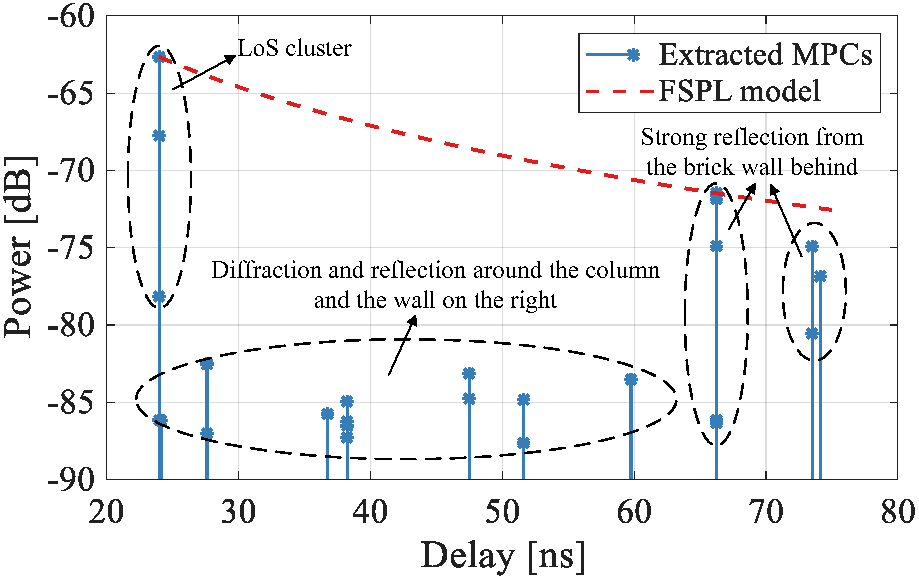}
\label{fig:PDP}}
\caption{The distribution of the extracted MPCs from the PADP at 318 GHz of Rx6 in the (a) joint delay-AOA-AOD domain and (b) PDP alongside the FSPL model.}
\label{PS}
\vspace{-6pt}
\end{figure}

Since the original angular resolution is limited by the discrete scanning step size, the antenna radiation pattern is used to interpolate the PADP in the angular domain, thereby improving the resolution to $1^{\circ}$. This approach is widely adopted in directional channel sounding measurements \cite{Kimmo2024Anal}. Similarly, the sinc function is used to interpolate the PADP in the time domain to enhance the delay resolution to 0.025\,ns. A threshold-based denoising procedure is then applied to the refined PADP, and effective MPCs are identified as local peaks \cite{Katsu2016Est}. Fig.~\ref{fig:PADP} illustrates the extracted MPCs in the joint delay-AoA-AoD domain for location Rx6, which exhibits the most complex multipath conditions among all Rx points at 318\,GHz. The power $p_i$, delay $\tau_i$, and angle $\{\varphi_{\text{t},i},\varphi_{\text{r},i}\}$ of each effective MPC can be directly obtained from the PADP.

\subsection{Multipath Component Clustering}
Distance-based MPC clustering is a widely used low-complexity single-snapshot-based method \cite{huang2022artificial}. In \cite{czink2006framework}, the multipath component distance (MCD) is used as the metric for measuring the distance between MPCs, and clustering is performed using the KPowerMeans algorithm. However, the performance of KPowerMeans is sensitive to initialization, necessitating multiple iterations of randomization of the initial parameters. To address this, we introduce an MCD-based pre-clustering initialization procedure prior to applying KPowerMeans.

The MCD in the Mahalanobis-distance metric \cite{chen2022framework} between the $i$-th MPC and the $j$-th MPC in PADP is defined as
\begin{equation}\label{equ:mcd}
\begin{split}
MCD&_{i,j}=\|[\Delta\tau_{i,j},\Delta\delta_{\text{t},i,j},\Delta\delta_{\text{r},i,j}]^T\|_{\textbf{A}}\\
&=\sqrt{[\Delta\tau_{i,j},\Delta\delta_{\text{t},i,j},\Delta\delta_{\text{r},i,j}]\textbf{A}[\Delta\tau_{i,j},\Delta\delta_{\text{t},i,j},\Delta\delta_{\text{r},i,j}]^T},
\end{split}
\end{equation}
where $\Delta\tau_{i,j}$ denotes the delay difference between the two MPCs, and $\Delta\bm{\delta}_{\text{t/r},i,j}=\hat{\bm{r}}_{\text{t/r},i}-\hat{\bm{r}}_{\text{t/r},j}$, with $\hat{\bm{r}}$ representing the unit direction vector. The $\hat{\bm{r}}_{\text{t/r},i}$ can be expressed as
\begin{equation}\label{equ:udv}
\hat{\bm{r}}_{\text{t/r},i}=\left[\begin{array}{c}
   \sin{\phi_{\text{t/r},i}}\cos{\varphi_{\text{t/r},i}}\\
   \sin{\phi_{\text{t/r},i}}\sin{\varphi_{\text{t/r},i}}\\
   \cos{\phi_{\text{t/r},i}}
\end{array}\right]^T.
\end{equation}
The matrix $\textbf{A}$ is a real semi-definite weighting matrix that balances the contributions of delay and direction terms. It is typically set to $\text{diag}(\xi^2,0.25,0.25,0.25,0.25,0.25,0.25)$, where the details of $\xi$ referred to \cite{steinbauer2002quantify}.

The initialization procedure consists of the following steps:
\begin{enumerate}
    \item Calculate the MCD for each pair of MPCs. A dynamic threshold $\Gamma=\|[\delta_{\tau},\bm{\delta}_{\text{t}},\bm{\delta}_{\text{r}}]\|_{\textbf{A}}$ is defined, where the elements represent allowable deviations in intra-cluster delay, Tx and Rx direction deviation, respectively:
    \begin{equation}\label{equ:thr}
    \left\{\begin{array}{l}
       \delta_{\tau}=0.2\cdot \sigma_{\tau}\\
       \bm{\delta}_{\text{t}}=[\sqrt{2-2\cos{(\sigma_{\varphi_{\text{t}}}/4)}},0,0]\\
       \bm{\delta}_{\text{r}}=[\sqrt{2-2\cos{(\sigma_{\varphi_{\text{r}}}/4)}},0,0]
    \end{array}\right..
    \end{equation}
    Here, $\sigma_{\bullet}$ denotes the standard deviation of the item at the given subscript across all measurements. A detailed derivation of these three parameters is provided in Appendix. 

    \item Select the MPC whose smallest MCD is the smallest to act as the reference target. From the remaining unclustered MPCs, identify those whose distance to the reference is below the threshold $\Gamma$ and rank among the top 3 smallest distances (a parameter that can be flexibly adjusted). Group these identified MPCs together with the reference MPC, and repeat this procedure iteratively to generate a set of pre-segmented clusters.

    \item Randomly select one MPC from each pre-segmented cluster in descending order of cluster power as an initial point until the required number of initial MPCs is reached. If the number of pre-segmented clusters is insufficient, supplement with the MPCs with the largest minimum MCD.
\end{enumerate}

Since there is no prior knowledge of the number of clusters in the channel, each interesting number must be evaluated. Therefore, the cluster validity index (CVI) is required to assess the clustering performance when selecting different numbers of clusters. Conventional CVIs, such as the Cali\~{n}ski-Harabasz index, Davies-Bouldin index, and Silhouette index, rely on intra-cluster compactness and inter-cluster separation \cite{czink2006framework, rousseeuw1987silhouettes}. However, these metrics are not directly applicable to THz channels because the compactness of a cluster with only one MPC cannot be evaluated. To address this limitation, we extend the Silhouette index. The original Silhouette value of the $i$-th MPC in the $k$-th cluster is given by
\begin{equation}\label{equ:ss}
S_{i,k}=\frac{d_{i,k}-b_{i,k}}{\max(d_{i,k},b_{i,k})},
\end{equation}
where $d_{i,k}$ is the minimum value of the averages of the MCDs between the $i$-th MPC in the $k$-th cluster and all MPCs within another cluster, and $b_{i,k}$ is the average of the MCDs between the $i$-th MPC in the $k$-th cluster and all other MPCs in the $k$-th cluster. For the $k'$-th cluster containing only one MPC, the Silhouette score is redefined as
\begin{equation}\label{equ:iss}
S_{1,k'}=\frac{d_{1,k'}-\max(b_{i,k},\Gamma_{\text{res}})}{\max(d_{1,k'},\max(b_{i,k},\Gamma_{\text{res}}))},
\end{equation}
where $\Gamma_{\text{res}}$ represents the MCD corresponding to two MPCs that are spaced apart in the temporal and spatial domains by their respective resolutions. This modification ensures robust clustering evaluation even in extreme numbers of clusters.
\begin{figure*}[!t]
\centering
\subfigure[][]{\includegraphics[width=2.35in]{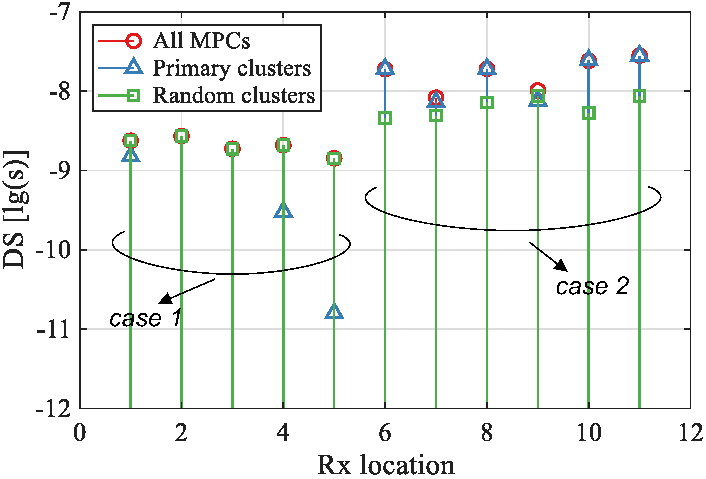}\label{Fig:DS_FH318}}
\subfigure[][]{\includegraphics[width=2.35in]{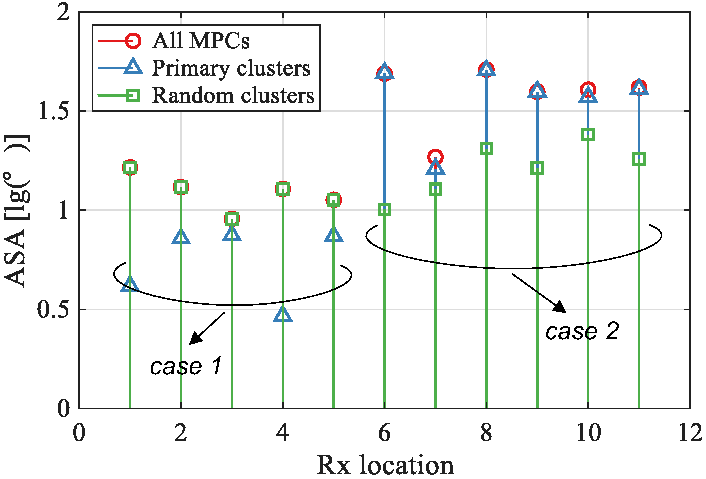}\label{Fig:ASA_FH318}}
\subfigure[][]{\includegraphics[width=2.35in]{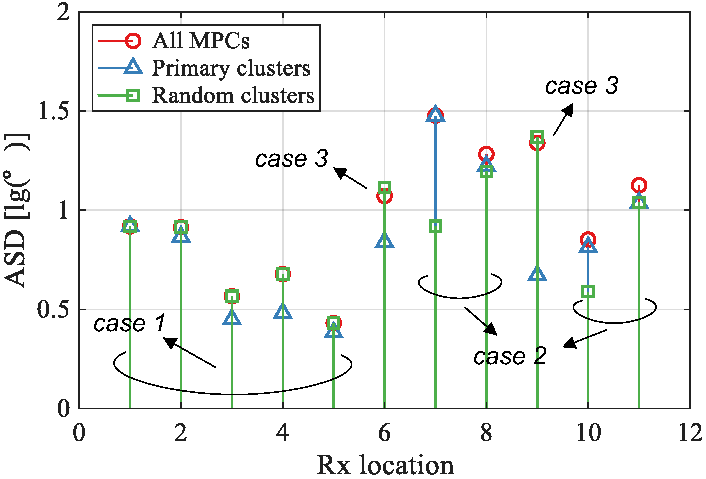}\label{Fig:ASD_FH318}}
\caption{Comparison of the (a) DS, (b) ASA, and (c) ASD of all primary clusters, all random clusters, and all MPCs at 318\,GHz.}
\label{sc}
\vspace{-6pt}
\end{figure*}

\subsection{Propagation Characteristics Analysis}
As discussed in Section \ref{sec:I}, THz channels exhibit not only sparsity but also quasi-optical propagation properties. Fig.~\ref{fig:PDP} illustrates the power delay profile (PDP) at location Rx6, along with the free-space path loss (FSPL) curve aligned with the LoS path. Strong non-LoS (NLoS) paths can be observed, and the delay and power of these paths and the LoS path closely follow the FSPL trend. In contrast, other NLoS paths exhibit no apparent trend between power and delay, indicating that the NLoS paths consistent with the FSPL model possess fundamentally different propagation characteristics from the remaining NLoS paths.

Based on this observation, clusters with power levels no more than 6\,dB below the FSPL curve are defined as primary clusters, while the remaining clusters are classified as random clusters. Fig.~\ref{sc} compares the delay spread (DS), azimuth angular spread of arrival (ASA), and azimuth angular spread of departure (ASD) for all clusters, all primary clusters, and all random clusters. To intuitively illustrate their contributions, the spread calculations for random clusters take the LoS cluster into account.

As illustrated in Fig.~\ref{sc}, the relationships between the delay and angular spreads of these three types of clusters can be categorized into three distinct cases: 

Case\,1: The overall spread closely aligns with the spread of random clusters. This indicates an absence of strong reflections, meaning the primary cluster consists exclusively of the LoS path.

Case\,2: The overall spread closely approximates the spread of primary clusters, demonstrating that strong reflections completely dominate the channel dispersion.

Case\,3: The overall spread aligns with neither primary nor random clusters, reflecting a mixed environment where both propagation phenomena contribute to the overall spread.

In practice, almost only the first two cases occur, meaning that either primary clusters are dominant or no primary clusters exist except for the LoS cluster. This observation can serve as the foundation for analyzing or even modeling primary clusters and random clusters separately.
\begin{table}[!t]
\caption{Channel parameters specifically for primary clusters and random clusters in an industrial environment at 318\,GHz}
\begin{center}
\begin{tabular}{|c|c|c|c|c|}
\hline
\multicolumn{2}{|c|}{\textbf{Parameter}} & \textbf{Unit} & \textbf{Primary} & \textbf{Random}\\
\hline
\multirow{2}{*}{Number of clusters} & $\mu$ & - & 2.0 & 3.9\\
\cline{2-5} 
& $\sigma$ & - & 1.2 & 2.9\\
\hline
Number of rays per cluster & $\mu$ & - & 4.6 & 1.8\\
\hline
Interarrival of clusters & $\mu$ & ns & 27.4 & 9.8\\
\hline
Power decay rate & $\Gamma$ & dB/ns & -0.173 & -0.026\\
\hline
Per cluster shadowing & $\xi$ & dB & 2.87 & 4.10\\
\hline
\multirow{2}{*}{AoA} & $\mu$ & deg & - & 18.6\\
\cline{2-5} 
& $\sigma$ & deg & - & 94.8\\
\hline
\multirow{2}{*}{AoD} & $\mu$ & deg & - & -32.9\\
\cline{2-5} 
& $\sigma$ & deg & - & 35.3\\
\hline
\end{tabular}
\label{sv}
\end{center}
\vspace{-6pt}
\end{table}

\section{Statistical Characteristics Analysis}
\label{sec:IV}
The necessity of differentiating between primary and random clusters has been established in the previous section. In this section, we present a statistical analysis of these two cluster types separately, focusing on key parameters relevant to channel modeling, including the number of clusters and rays, delay, power, and angular distributions. Unlike conventional analysis that primarily emphasizes large-scale and small-scale parameters, this comparative analysis aims to highlight the differences between primary and random clusters in THz channels and provide a reference for THz cluster-based channel modeling. All statistical parameters are summarized in Table~\ref{sv}.

\subsection{Number of Clusters and Rays}
\begin{figure}[!t]
\centering
\subfigure[]{\includegraphics[width=3.0in]{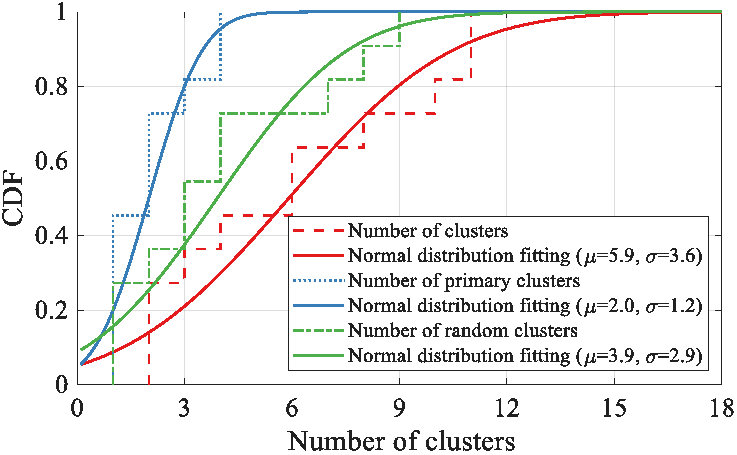}
\label{fig:NoC}} 
\subfigure[]{\includegraphics[width=3.0in]{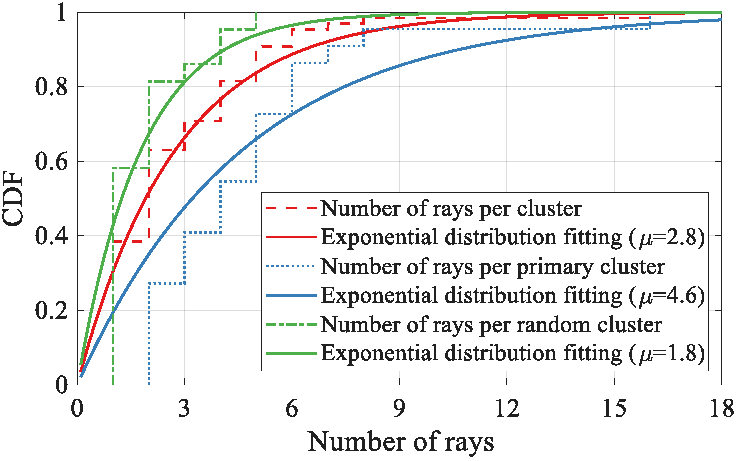}
\label{fig:NoR}}
\caption{Comparison of empirical and fitting CDFs of the number of (a) primary and random clusters and (b) rays per primary and random cluster.}
\label{Number}
\vspace{-6pt}
\end{figure}

The number of clusters reflects both the richness of scatterers and the attenuation of the propagation channel. Fig.~\ref{fig:NoC} shows the cumulative distribution functions (CDFs) of the number of primary and random clusters. The average total number of clusters is 5.9, which is slightly higher than the value of 3.4 at 142\,GHz reported in \cite{ju2024statistical}. This discrepancy may arise from differences in clustering methodologies or from the richer multipath conditions present in the factory environment in this study. However, the average number of primary clusters is only 2, with a noticeably smaller variance compared to that of random clusters.

Unlike the number of clusters, the number of rays per cluster in 3GPP is set to a relatively large number to form a specific shape in the angular domain \cite{kyosti2012channel}. In practice, however, the observed number of rays per cluster depends not only on the scattering environment but also on the angular and delay resolutions of the measurement system and the MPC extraction algorithm. Therefore, the numerical values provided here are primarily intended for comparison analysis. As shown in Fig.~\ref{fig:NoR}, primary clusters tend to contain slightly more rays than random clusters. This can be attributed to their higher power levels, which result in broader observable power spectra within the dynamic range of the channel sounder, making multiple rays more detectable at the Rx.

\subsection{Delay and Power}
\begin{figure}[!t]
\centerline{\includegraphics[width=3.0in]{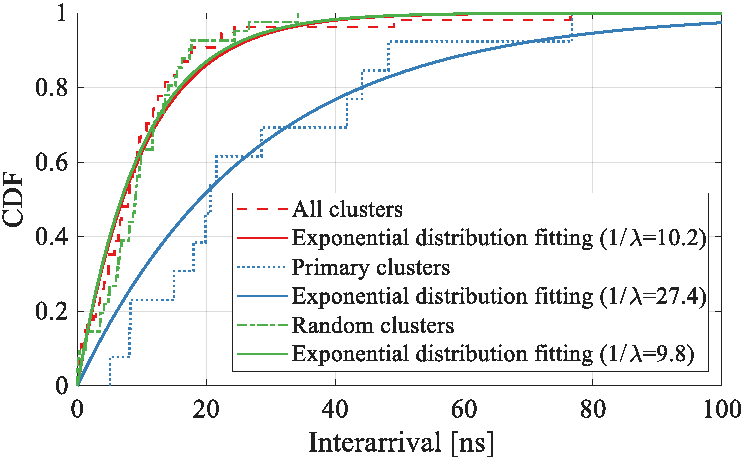}}
\caption{Comparison of empirical and fitting CDFs of the interarrival time of primary and random clusters.}
\label{Interarrival}
\vspace{-6pt}
\end{figure}

The cluster arrivals in the 3GPP TR38.901 model for 0.5-100\,GHz are modeled as a Poisson process \cite{3GPP38901}. Fig.~\ref{Interarrival} presents the CDFs of the interarrival times for primary and random clusters, along with their exponential distribution fits. All fits passed the Kolmogorov–Smirnov (K–S) test at the 5\% significance level. The mean interarrival time of random clusters is comparable to that of the overall clusters, while the mean interarrival time of primary clusters is larger than the former two.

\begin{figure}[!t]
\centerline{\includegraphics[width=3.0in]{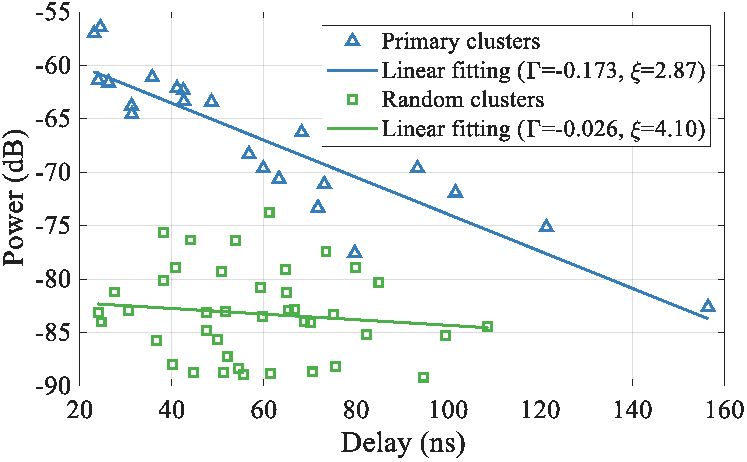}}
\caption{Comparison of empirical and fitting results of power decay with delay of primary and random clusters.}
\label{PvD}
\vspace{-6pt}
\end{figure}

The S-V model has been verified to remain applicable to THz channels \cite{saleh1987statistical}. Fig.~\ref{PvD} illustrates the relationship between cluster power and delay, along with the linear regression results. The linear fit for primary clusters yielded a lower root mean square error than that for random clusters (2.71 vs. 4.26). Evidently, primary clusters exhibit a more severe temporal decay than random clusters. A reasonable explanation is that primary clusters commonly experience fewer specular reflections and suffer lower loss, while random clusters experience more diffusion or diffraction with greater loss, leading to increased propagation loss and more irregular power variations over delay. This observation is further supported by the per cluster shadowing values, where the primary clusters' value of 2.87\,dB is lower than the 4.10\,dB of random clusters.

\subsection{Angular Distribution}
\begin{figure}[!t]
\centerline{\includegraphics[width=3.0in]{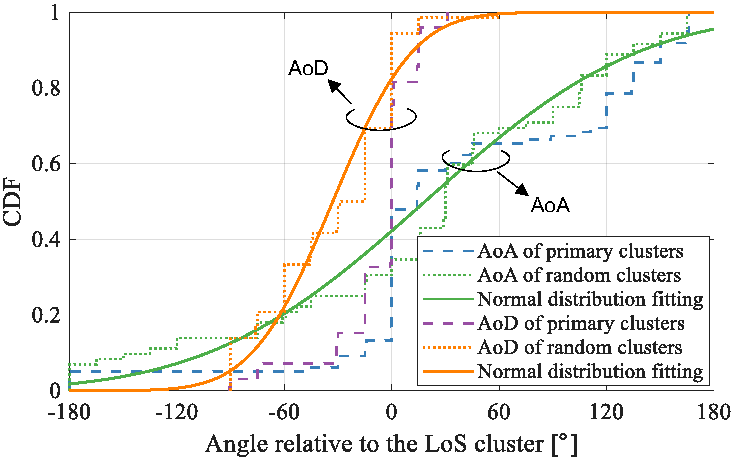}}
\caption{Comparison of empirical and fitting CDFs of the AoA and AoD of primary and random clusters.}
\label{Angle}
\vspace{-6pt}
\end{figure}

The angular distribution depends on both the Tx–Rx geometry and the environment. In LoS scenarios, channel modeling focuses on angular deviations relative to the LoS direction. The distributions of AoA and AoD of primary and random clusters are shown in Fig.~\ref{Angle}. Due to the limited angular scanning range of the Tx antenna, the AoD distribution is inherently biased and truncated. Despite this limitation, the AoD of random clusters approximately follows a normal distribution $\mathcal{N}(-32.9,35.3^2)$, while the AoA follows $\mathcal{N}(18.6,94.8^2)$. In contrast, the AoA and AoD of primary clusters appear to follow no discernible distribution pattern, suggesting a considerable dependence on specific environmental conditions.

\section{Conclusions}
\label{sec:V}
In this paper, we investigated the propagation characteristics of 318\,GHz THz channels in a typical industrial environment. Driven by the observed sparsity and the prominent presence of strong reflections, we proposed a novel MCD-based initialization procedure for the KPowerMeans clustering, alongside an extended Silhouette index to better evaluate sparse multipath components. By introducing a threshold based on the free-space propagation model, we successfully distinguished primary clusters (strong reflections) from random clusters. Our subsequent statistical analysis validated this classification, revealing significant differences in their respective channel spread characteristics, power decay rates, and angular distributions. This stark contrast underscores the necessity of modeling primary and random clusters separately to accurately capture THz channel dynamics. Ultimately, these findings provide valuable insights and practical guidance for developing robust, high-fidelity THz stochastic channel models for future IIoT deployments. To further generalize these results, extensive measurement campaigns across diverse industrial scenarios remain an important area for future work.

\section*{Appendix\\
Dynamic Threshold in the Initialization Process}
\label{sec:Appendix}
Assume that a channel consists of $M$ clusters that are mutually orthogonal and exhibit similar intra-cluster characteristics. The parameter range of MPCs within a single cluster should not exceed $1/M$ of the overall parameter range across all MPCs in the channel. Based on the empirical number of clusters and the fact that the threshold should not be set too low, $M$ is set to 5 in this work.

According to the generation procedure for the delays in \cite{3GPP38901}, the threshold for delay corresponding to the 20$\%$ point of the delay distribution can be expressed as 
\begin{equation}\label{equ:t_delay}
\delta_{\tau}=0.2\cdot \text{DS}\cdot r_{\tau}=0.2\cdot \sigma_{\tau}.
\end{equation}
The azimuth and elevation angles are modeled as wrapped Gaussian and Laplacian distributions, respectively, in \cite{3GPP38901}. Therefore, the thresholds for azimuth and elevation angles corresponding to the 20$\%$ point of their distributions are given by $\varphi'=\sigma_{\varphi}/4$ and $\phi'=\sigma_{\phi}/4.5$, respectively. The squared Euclidean distance between two unit direction vectors with azimuth difference $\varphi'$ and elevation difference $\phi'$ can be calculated as
\begin{equation}\label{equ:t_angle}
\begin{split}
&\Delta\delta\bm{I}_3\Delta\delta^T\\
&=2-2\cos\phi\cos(\phi\pm\phi')-2\sin\phi\sin(\phi\pm\phi')\cos\varphi'\\
&\leq2-2(\cos\phi\cos(\phi\pm\phi')-2\sin\phi\sin(\phi\pm\phi'))\cos\varphi'\\
&=2-2\cos\phi'\cos\varphi',
\end{split}
\end{equation}
where $\bm{I}_3$ denotes the $3\times3$ identity matrix and $\phi$ represents an arbitrary elevation angle. Since elevation-domain scanning is not performed in this work, the threshold for the angle is simplified as $[\sqrt{2-2\cos{(\sigma_{\varphi}/4)}},0,0]$.

\section*{Acknowledgment}
This work was partly supported by the UK Engineering and Physical Sciences Research Council (EPSRC) through the EPSRC Hub on All Spectrum Connectivity (EP/X040569/1 and EP/Y037197/1). The work of Pekka Ky\"osti was supported by the Research Council of Finland (former Academy of Finland) project (Grant no. 348980) and 6G Flagship Programme (Grant no. 346208). Keysight Technologies, Inc. has supported the research with measurement equipment donation.

\balance
\bibliographystyle{IEEEtran}
\bibliography{ref}

\end{document}